\documentclass{ws-ijmpa}

\begin{document}

\markboth{Elena C\'aceres}
{Glueball Spectrum and Regge Trajectory from Supergravity}

%
%

\title{GLUEBALL SPECTRUM AND REGGE TRAJECTORY FROM SUPERGRAVITY}

\author{ELENA CACERES\footnote{caceres@fis.cinvestav.mx}}

\address{Department of Physics, CINVESTAV\\Av. Polit\'ecnico Nac. 2508,Mexico D.F. 73600\\Mexico
}



\maketitle


\begin{abstract}
Brief review of the status of the glueball spectrum in the deformed 
conifold background. Talk based on work done in collaboration with with R. Her\'andez and X. Amador. 
\end{abstract}

\section{Introduction}	
The formulation of a gauge/string duality \cite{Maldacena}
provides a new framework to study confining phenomena. The duality was originally formulated between type IIB string theory propagating in Anti de Sitter (AdS)  space and a conformal field theory (CFT).  Nowadays several
supergravity
duals to confining gauge theories are known \cite{PolchinskiStrassler,KlebanovStrassler}. Some of this backgrounds are deformations of the original AdS/CFT duality. This is the case of  AdS blackhole backgrounds dual to a finite temperature field theory. The spectrum in these backgrounds has been thoroughly studied by Brower, Mathur and Tan\cite{BrowerMathurTan}. 
In the present work we focus on the Klebanov-Strassler (KS) IIB supergravity solution
\cite{KlebanovStrassler}. This background  describes N regular and M fractional
D3 branes on a deformed conifold space; it is conjectured to be  dual to
a cascading gauge theory with $SU(M+N)\times SU(N)$ gauge group in
the ultraviolet and flows to $SU(M)$ Super-Yang Mills in the infrared. The
glueball spectrum for $0^{++}$ and $1^{--}$ in the KS background
was obtained in Ref.\refcite{CaceresHernandez} and  the spin 2 glueball mass was recently obtained in Ref.\refcite{AmadorCaceres}. We will review this results and address possible comparisons with lattice results. In particular, features of the Regge trajectory as obtained from this spectrum.

Regge theory successfully describes a large quantity of experimental data. It predicts that
composite particles of a given set of quantum numbers, different only in their spin, will lie on a linear trajectory \[
J=\alpha_{0}+\alpha^{'}t\] where $J$ is the spin and $t$ is the mass squared. Regge theory treats the strong interaction as the exchange
of a complete trajectory of particles. With the inclusion of a soft pomeron, this approach successfully describes the high energy
scattering of hadrons. It is thus interesting to explore Regge trajectories from the gauge/ gravity duality. We show that the KS background
predicts a \textit{linear} glueball trajectory, $J=-2.29+0.23\,\, t$. Unlike the scenario where glueball masses are identified with
classical solutions of folded strings, here there is no \textit{a priori} reason for the Regge trajectory to be linear. That it turns
out to be so is remarkable.


\section{The Deformed Conifold}

The Klebanov-Strassler background  \cite{KlebanovStrassler}   
 is a solution to type II B supergravity equations with
non-zero three  and five forms  and a constant dilaton.
%
The KS solution is rich in
interesting physical phenomena; exhibits confinement, chiral symmetry
breaking, dimensional transmutation, domain walls etc. We  will
review some aspects of the KS background necessary for the next sections
. 

The solution describes N regular and M fractional D3 branes on a deformed conifold and is given by,
\begin{equation}
ds_{10}^{2}=h^{-1/2}(\tau)dx_{n}dx_{n}+h^{1/2}(\tau)ds_{6}^{2}\label{eq:tendimspace}\end{equation}
 where \[
h(\tau)=(g_{s}M\alpha')^{2}2^{2/3}\epsilon^{-8/3}I(\tau)\equiv\int_{\tau}^{\infty}dx\ \frac{(x\coth x-1)(\sinh(2x)-2x)^{1/3}}{\sinh^{2}x},\]

$$ {\cal F}_5 = B_2\wedge F_3 = {g_s M^2 (\alpha')^2\over 4} l(\tau) g^1\wedge g^2\wedge g^3\wedge g^4\wedge g^5 $$ 

$$ F_3 ={M\alpha'\over 2} \left \{g^5\wedge g^3\wedge g^4 + d [ F(\tau) (g^1\wedge g^3 + g^2\wedge g^4)]\right \}  $$

 $$H_3 = dB_2 = {g_s M \alpha'\over 2} \bigg[ d\tau\wedge (f' g^1\wedge g^2 +  k' g^3\wedge g^4)  + {1\over 2} (k-f) g^5\wedge (g^1\wedge g^3 + g^2\wedge g^4)\bigg].\nonumber
  $$
The precise form of the functions  defining the deformed conifold metric, $ds_6^2$,  the three-form, $F_3$, and five-form,$F_5$,  can be found in Ref.\refcite{KlebanovStrassler}.

In the infrared region , $\tau\rightarrow0$,  the metric becomes
\begin{eqnarray} ds_{10}^2  &\rightarrow&  { \varepsilon^{4/3}\over 2^{1/3} a_0^{1/2} g_s M\alpha'} dx_n dx_n  + a_0^{1/2} 6^{-1/3} (g_s M\alpha') \bigg \{ {1\over 2} d\tau^2  + {1\over 2} (g^5)^2 \nonumber \\ && + (g^3)^2 + (g^4)^2      + {1\over 4}\tau^2 [(g^1)^2 + (g^2)^2] \bigg \} \ . \label{apex} \end{eqnarray} 
The parameter $\epsilon^{2/3}$ has
dimensions of length and measures the deformation of the conifold.
From the IR metric (\ref{apex}) we see  that $\epsilon^{2/3}$ also
sets the dynamically generated 4-d mass scale $\frac{\epsilon^{2/3}}{\alpha'\sqrt{g_{s}M}}$ and the
glueball masses scale  $m_{glueball}\sim\frac{{\epsilon^{2/3}}}{\alpha^{'}g_{s}M}$.

\section{Glueball Masses and Regge Trajectory}
\subsection{$O^{++}$ and $1^{--}$}
The simplest case to consider is the $0^{++}$ which is dual to  fluctuations of the dilaton field $\tilde{\phi}$. After linearizing the equations of motion it is
standard procedure to make an expansion in harmonics on the angular
part of the transverse space, $\tilde{\phi}=\Sigma_I \Phi^I(\bar x, \tau) Y^I (\bar\theta)$, where $\bar x$ is shorthand for $x_1,x_2,x_3,x_4$  coordinates in Minkowski space,  $\tau$ ir a radial coordinate and $\bar \theta $ denotes all the angular coordinates.  Collecting the equations for scalar harmonics we find that the dilaton fluctuation decouples from the other fluctuations, the relevant equation is $$\square {\tilde \Phi}(\bar x,\tau)=0.$$ The next step is to expand  in plane waves $\tilde{\phi}(\tau,{\bar x})=e^{i{\bf k}\cdot {\bar x}}f(\tau)$. Using the ten dimensional metric (\ref{eq:tendimspace})  we get,
\begin{equation}
3.2^{1/3} \frac {d}{d \tau} \Big[ (\sinh (2 \tau) - 2 \tau)^{2/3} 
\frac {df(\tau)}{d \tau} \Big] - (k^2 \epsilon^{4/3}) \sinh^2 (\tau) h(\tau) f((\tau) = 0.
\label{b}
\end{equation}
where each mode has mass $m^2 =-k^2$. The spin zero glueball masses are identified with the values of $k^2$ for which 
eq.(\ref{b}) has a normalizable solution. Solving this eigenvalue problem numerically we get,
$$m^2(0^{++})=9.78,\quad \quad \quad m^2(0^{++*})=33.17$$
where the masses are measured in units of the conifold deformation $\epsilon ^{\frac{4}{3}}$.The same procedure can be applied for the $1^{--}$ that is identified with a gauge field in the supergravity side. In this case the equations do not decouple and we have to solve a system of equations
(see Ref. \refcite{CaceresHernandez} for details). We obtain, 
$m^2(1^{--})=14.05$,\quad $m^2(1^{--*})=42.90$.
\subsection{Spin Two Glueball}
Finding the  spin two glueball mass is,technically, more challenging since we have to solve for fluctuations of the metric. Consider the metric, 
\[
g_{MN}=g_{MN}^{KS}+h_{MN}\]
 where $g_{MN}^{KS}$ is the Klebanov-Strassler background metric
(eq. \ref{eq:tendimspace}) and $h_{MN}$ denotes fluctuations around
this background.  In the present case we are interested
in infrared phenomena \textit{i.e.} in the $\tau\rightarrow0$ region.
In this region the angular part of the deformed conifold behaves as
an $S^{3}$ that remains finite -with radius of order $g_{s}M$ at
$\tau=0-$ and an $S^{2}$ which shrinks like $\tau^{2}$ . Thus,
for small $\tau$ it is appropriate to expand in spherical harmonics
on the $S^{3}$. 

After introducing the expansion in harmonics in the IIB supergravity
equations and keeping in mind that we are interested in fluctuations
on the four dimensional space transverse to the deformed conifold
we find that the linearized equation for the fluctuations is\begin{eqnarray}
-\frac{1}{2}\nabla^{\lambda}\nabla_{\lambda}h_{ij}(\tau,\bar{x})\,+\frac{1}{2}\nabla^{l}\nabla_{i}h_{lj}(\tau,\bar{x})\,+\frac{1}{2}\nabla^{l}\nabla_{j}h_{li}(\tau,\bar{x})=\,\,\,\,\,\,\,\,\,\,\,\,\,\,\,\,\,\,\,\,\,\,\,\,\,\,\,\,\,\,\,\nonumber \\
\left(\,\frac{g_{s}^{2}}{96}\,\left(\frac{1}{5}\star\mathcal{F}_{5}^{KS}\cdot\star\mathcal{F}_{5}^{KS}\right)\,-\left.\frac{g_{s}^{2}}{48}H_{3}^{KS}\cdot H_{3}^{KS}-\frac{1}{48}F_{3}^{KS}\cdot F_{3}^{KS}\right)h_{ij}(\tau,\bar{x})\right.\label{eq:einsteinlinear}\\
\nonumber \end{eqnarray}
 where the covariant derivative is with respect to the full KS background
(\ref{eq:tendimspace}). Expanding in plane waves, \[
g_{ij}(\tau,\bar{x})=h_{ij}(\tau)e^{i{\bf k}x},\]
a mode of momentum $k$ has a mass ${\bf k}^{2}=-m^{2}.$ The spin 2 representation
of the $SO(3)$ symmetry in $x_{2},x_{3},x_{4}$ is a symmetric traceless
tensor. Choosing a gauge $g_{1i}(\tau,\bar{x})=0$, the fluctuation
$h_{ij}(\tau)$ ( $i,j=1,2,3,4$) has five independent components
corresponding to the five polarizations of the $2^{++}.$ As expected,
all five satisfy the same equation of motion and are thus degenerate.
Denoting $g(\tau)\equiv h_{22}(\tau)=h_{33}(\tau)=h_{23}(\tau)=h_{24}(\tau)=h_{34}(\tau)$
we obtain (see Appendix for details) from (\ref{eq:einsteinlinear}),
\begin{equation}
\frac{d^{2}}{d\tau^{2}}g\left(\tau\right)+A\left(\tau\right)\frac{d}{d\tau}g\left(\tau\right)+\left(B\left(\tau\right)-\frac{g_{s}^{2}\alpha^{2}M^{2}}{2^{1/3}\epsilon^{4/3}}I(\tau)G_{55}(\tau){\bf k}^{2}\right)g\left(\tau\right)=0\label{eq:sugraeq}\end{equation}
where, \[
A(\tau)=\frac{d\ln(G_{99}(\tau))}{d\tau}+\frac{d\ln(G_{77}(\tau))}{d\tau}+\frac{d\ln I(\tau)}{d\tau}\]
and
\begin{eqnarray*}
&&B(\tau) =  \frac{-2^{(1/3)}(1-F(\tau))^{2}}{8I(\tau)G_{77}(\tau)^{2}}-(\frac{dk(\tau)}{d\tau})^{2}-\frac{2^{1/3}(k(\tau)-f(\tau))^{2}}{16G_{77}(\tau)G_{99}(\tau)} +
4(\frac{dF(\tau)}{d\tau})^{2} -\\ &&\frac{2^{1/3}}{8I(\tau)G_{99}(\tau)^{2}}((\frac{df(\tau)}{d\tau})^{2}+F(\tau)^{2}) +\frac{1}{4I(\tau)^{2}}\left(\frac{dI(\tau)}{d\tau}\right)^{2}-\frac{2^{5/3}l(\tau)^{2}}{I(\tau)^{2}K(\tau)^{4}(\cosh^{2}\tau-1)^{2}}\\
\end{eqnarray*}
$G_{77}(\tau)$ and $G_{99}(\tau)$
are redefinitions of the background metric $g_{MN}^{KS}(\tau)$ such
that $g_{ii}^{KS}(\tau)=h(\tau)^{-1/2}G_{ii}(\tau)$ for $i=1...4$
and $g_{\mu\mu}^{KS}(\tau)=h(\tau)^{1/2}\epsilon^{4/3}G_{\mu\mu}(\tau)$
for $\mu=5...10$ , they do not contain dimensionfull quantities. 
Spin two glueball masses are identified with the values of ${\bf k}^2 $ for which there is a solution of (\ref{eq:sugraeq}) with  boundary conditions  that will ensure the solution is renormalizable. 
\subsection{Results}
A spreviously stated, equations (\ref{b}) and  (\ref{eq:sugraeq}) are  eigenvalue problems. They can be
solved exactly by a variety of numerical methods. The boundary conditions
at infinity are  found by demanding normalizability of the states.  We used a  "shooting technique" to find the eigenvalues. This method is very
accurate for low lying states but requires an initial guess for the
eigenvalue. We used as initial guess the value found from a  WKB approximation. Equation (\ref{eq:sugraeq})
is particulaarly sensitive to accumulation of numerical error due to the combination of hyperbolic functions involved in the coefficients.
In order to overcome this difficulty 
we calculate  $A(\tau)$and $B(\tau)$ with 20 digits of precision .
With this technique we  find a very stable eigenvalue for  $m^{2}(2^{++})=18.33$
We collect  the results for the lowest lying $0^{++}$, $1^{--}$ and $2^{++}$ in Table 1.
\begin{table}[h]
\tbl{Glueball masses}
{\begin{tabular}{@{}c|ccc@{}} \hline
State &\ \ $0^{++}$ & $1^{--}$&$2^{++}$\\
\hline
{$Mass^{2}/\epsilon^{4/3}$}&\ \ 9.78 &14.05&18.33\\
\hline
\end{tabular}}
\end{table}
The Chew-Frautschi plot for the glueball trajectory obtained with
these values is shown in Figure 2. It is remarkable that the three
states lie on a straight line. For large quantum numbers it is known
that glueballs can be identified with spinning folded closed strings.
In that approach a linear Regge trajectory is no surprise since it
is built in the formalism. But in the present framework, where we
identify masses with eigenvalues of equations of motion, there is
no \textit{a priori} reason for the eigenvalues to lie on a straight
line. The fact that it is so is remarkable. The glueball Regge trajectory
obtained from the KS model is $J=-2.2+0.23t $. 
\begin{figure}[h]
\centerline{\psfig{file=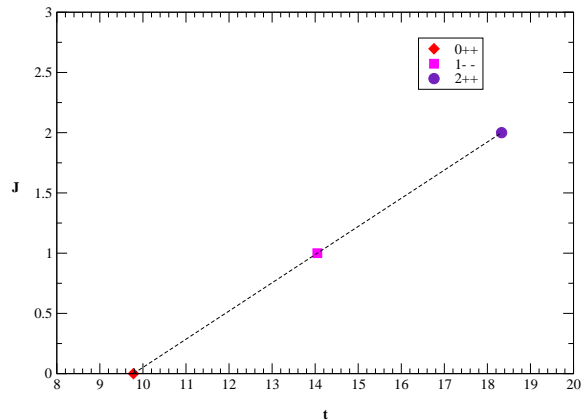,keepaspectratio,
  angle=-90,width=3.5in}}
\vspace*{8pt}
\caption{Glueball trajectory. The mass squared, $t$, is measured in units
of $\frac{\epsilon^{4/3}}{g_{s}^{2}\alpha'^{2}M^{2}}$.}
\end{figure}
We also  compare glueball mass ratios with lattice results. Lattice
data for $D=4$ is not as abundant as for $D=3$. Lucini and Teper
explored the $N\rightarrow\infty$ limit of $SU(N)$ Yang-Mills theory
in four dimensions \cite{LuciniTeper}. We present their results
in Table 2 and show that the agreement with supergravity results obtained
from the KS background is within one standard deviation. It is interesting
to include in the comparison lattice results for $SU(3)$  in $D=4$ \cite{MorningstarPeardon1}. 
\begin{table}[h]
\tbl{Glueball mass ratios calculated in the KS model and in two lattice simulations; $SU(\infty)$ and $SU(3)$.}
{\begin{tabular}{@{}cccc@{}} \toprule
 &KS model & Lattice $SU(\infty)$ & Lattice $QCD_{4}$\\
\hline
$m(0^{++*})/m(0^{++})$ & $1.84$ & $1.91(17)$ & $1.79(6)$ \footnotemark\\
\hline
$m(2^{++})/m(0^{++})$ & $1.37$ & $1.46(11)$ & $1.39(4)$\\ \botrule
\end{tabular}}
\end{table}
\addtocounter{footnote}{-1}

\footnotetext{a) Preliminary result, Ref.\refcite{MorningstarPeardon2}}

Note that the trajectory found here using a supergravity approach  does not reproduce what is expected 
for a Pomeron trajectory. In our opinion it is  too premature to make direct comparisons with experiimantal values for the Pomeron Regge trajectory. We consider that understanding Regge trajectories from supergravity is an important issue; the good agreement of mass ratios with lattice results and the fact that the trajectory obtained is linear show that we are making progress in that direction.  
\section*{Acknowledgments}
I am grateful to  Rafael Hern\'andez and Xavier Amador for 
enjoyable collaborations. This research was partially supported by the Mexican Council for Science and Technology (CONACYT).  
\flushright{{\it Dedicated to the memory of Iciar Isusi\\
\small Tarma, 1967 - Lima, 2002, Per\'u.}}

\end{document}